\documentclass{PoS}
\usepackage{cite}
\usepackage{subcaption}
\usepackage{booktabs,multirow,tabularx}

\newcommand\as{\alpha_{\mathrm{S}}} 
\newcommand\f[2]{\frac{#1}{#2}} 

\title{Standard Model Theory for Collider Physics}

\ShortTitle{SM Theory for Collider Physics}

\author{\speaker{Massimiliano Grazzini}\thanks{On leave of absence from INFN, Sezione di Firenze, Sesto Fiorentino, Florence, Italy.}\\
        Physics Department, University of Zurich, CH 8057 Zurich, Switzerland\\
        E-mail: \email{grazzini@physik.uzh.ch}}

\abstract{I briefly review the current theoretical description of Standard Model processes relevant for the LHC, and the tools that are used in the corresponding phenomenological applications. I discuss in particular the recent theoretical progress in NLO and NNLO QCD calculations, electroweak corrections, resummations, Monte Carlo tools and parton distribution functions. }

\FullConference{The European Physical Society Conference on High Energy Physics\\
		22--29 July 2015\\
		Vienna, Austria}

\begin{document}

\section{Introduction}

The first run of the Large Hadron Collider was a great success for the Standard Model (SM). Despite some intriguing hints of possible new physics signals, which definitely require more data from the Run 2 to be confirmed or disproved,
no evidence for new physics has been observed,
and the newly discovered Higgs resonance \cite{Aad:2012tfa,Chatrchyan:2012ufa}
appear very close to what expected in the SM. Accurate theoretical
predictions for SM processes were maybe not essential for the Higgs discovery,
but are crucial to interpret the Higgs signal. Searches for new physics in
the tail of kinematical distributions require a good control of SM backgrounds.

Our capability of predicting signals and backgrounds in high-energy hadron collisions is based on the QCD factorization theorem, according to which
the cross section for the production
of a hard final state $F$ can be written as
\begin{equation}
\label{ftheorem}
d\sigma^F=\sum_{a,b}\int dx_1dx_2\, f_{a/p}(x_1,\mu_F^2)f_{b/p}(x_2,\mu_F^2)\,d{\hat \sigma^F}_{ab}(x_1p_1,x_2p_2,\as(\mu_R^2),Q^2/\mu_F^2,Q^2/\mu_R^2)\, ,
\end{equation}
where the equality holds up to power suppressed terms ${\cal O}\left(\left(\Lambda/Q\right)^p\right)$.
The hard trigger $F$ can be for example a high-mass lepton pair, a high-$p_T$ lepton or jet, the mass of a heavy-quark, or of some new resonance. In all cases, the system is characterized
by a hard scale $Q$ which is much larger than the QCD scale $\Lambda$.
The functions $f_{a/p}(x,\mu_F^2)$ are the Parton Distribution Functions (PDFs), which express the probability of finding the parton $a$ in the proton at the factorization scale $\mu_F$. These functions cannot be perturbatively computed but must be extracted from global fits to the available data. The quantity $d{\hat \sigma^F}_{ab}$ is instead the partonic cross section, which can be perturbatively computed as a series expansion in the QCD coupling $\as(\mu_R)$,
but depends on the process we are considering.
From Eq.~(\ref{ftheorem}) it follows that, to obtain accurate theoretical predictions for the process at hand, we must rely both on reliable PDFs and on accurate
computations of the partonic cross sections. In the following we
discuss these issues in turn.

\section{PDFs and $\as$}
\label{sec:pdfs}

Traditionally, PDFs are parametrized at a low scale $Q_0=1-4$ GeV as
\begin{equation}
xf_{a/p}(x,Q_0^2)=x^{a_1} (1-x)^{a_2}P_a(x)
\end{equation}
where $P_a(x)$ is a slowly varying function.
After imposing the momentum sum rule the partons are then evolved to the desired $Q^2$ through DGLAP evolution
equations and then used to compute observables and fit the data.
The fits are usually based on a large number of data sets from fixed-target
and collider (HERA, Tevatron, LHC) experiments.

Recent PDF sets include MMHT14 \cite{Harland-Lang:2014zoa}, CT14 \cite{Dulat:2015mca}, NNPDF3.0 \cite{Nocera:2014gqa}, ABM12 \cite{Alekhin:2013nda}, HeraPDF2.0 \cite{Abramowicz:2015mha} and JR14 \cite{Jimenez-Delgado:2014twa}.
Overall the various PDF sets show reasonably good agreement,
but there are differences due
to the inclusion of different data sets in the fits, different theoretical assumptions, different treatment of heavy quarks, etc.

\begin{figure}[h]
\begin{center}
\includegraphics[width=0.6\textwidth]{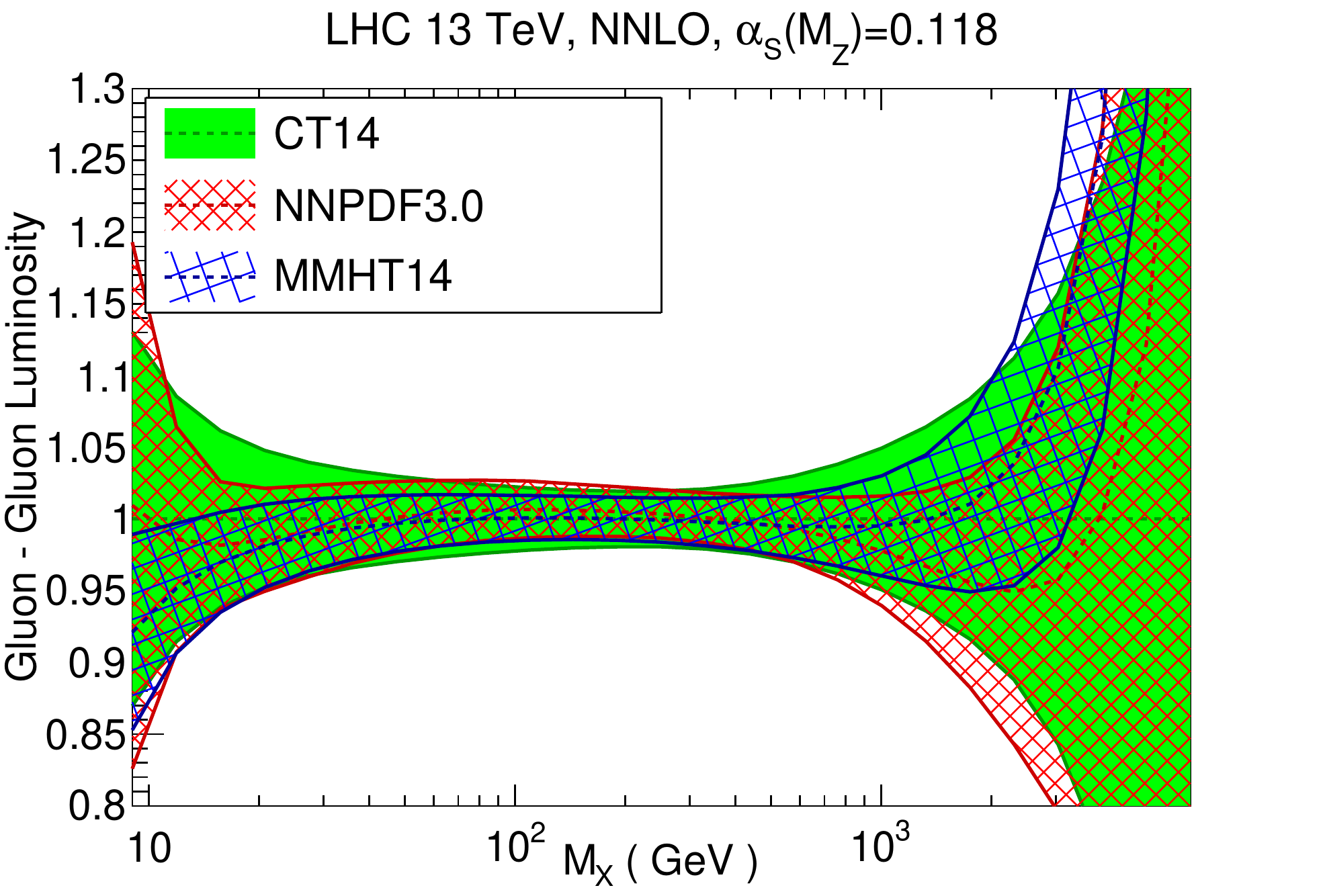}
\caption{Comparison of $gg$ luminosities from MMHT14, CT14 and NNPDF3.0.}
\label{fig:gg}
\end{center}
\end{figure}

In Fig.~\ref{fig:gg} (from Ref.~ \cite{Butterworth:2015oua}) we present a comparison of the gluon luminosities from the three fits which include
the widest range of experimental data: MMHT14, CT14 and NNPDF3.0.
Comparing to the situation with previous generation PDFs,
the overall consistency is significantly improved, with uncertainties down to the few percent level in the region $M_X\sim 100$ GeV,
and this is good news for Higgs boson production. This improvement arises from
a combination of methodological advancements and new experimental constraints in the fits.

The HeraPDF fit is based entirely on HERA data and, as such, has typically larger uncertainties. The ABM12 fit is based on DIS, Drell-Yan data, plus $t{\bar t}$ production at the LHC. The JR14 fit instead is based on a dynamical (valence-like) parametrization of the partons at low scale.

One important point is the treatment of hadron collider jet data.
Such data provide an important constraint on the gluon distribution at moderate and high values of $x$.
For this reason jet data are included in the MMHT14, CT14 and NNPDF3.0 fits although
the corresponding predictions are only available at next-to-leading order (NLO). An effort towards
the next-to-next-to-leading order (NNLO) computation is ongoing (see Sec.~\ref{sec:nnlo}).
MMHT includes approximated NNLO effects obtained from threshold resummation \cite{Kidonakis:2000gi}. This result, however, assumes that jets are massless in the threshold limit \cite{deFlorian:2013qia}. CT14 includes only jet data at the NLO. A different approach is adopted by NNPDF, which, using partial NNLO results, includes jet data only where the threshold approximation is expected to be valid \cite{Carrazza:2014hra}. 

A delicate and relevant
issue is the correlation of the PDFs with $\as(m_Z)$. The QCD coupling can indeed be fitted together with the PDFs, but it is more precisely obtained through
dedicated extractions from QCD studies at $e^+e^-$ colliders, $\tau$ decays, EW fits, etc. The current world average is \cite{Agashe:2014kda}
\begin{equation}
\as(m_Z)=0.1185\pm 0.0006
\end{equation}
with an uncertainty of $0.5\%$.
The value of $\as(m_Z)$ is crucial in many QCD predictions. An obvious example is the Higgs production cross section through gluon fusion, which starts at ${\cal O}(\as^2)$. Since this is a process with large perturbative corrections, by naively studying the effect of $\as$ variations one obtains that a $1\%$ increase in $\as(m_Z)$ may lead to
a $3\%$ increase in the Higgs cross section. A more careful study \cite{Harland-Lang:2015nxa}, which takes into account the correlations with the gluon density, finds a $1.5\%$ effect.

In any case, the central value of $\as(m_Z)$ and its uncertainty are clearly
a crucial input.
We note that a recent reassessment of the lattice result \cite{Aoki:2013ldr} leads to $\as(m_Z)=0.1184\pm 0.0012$, with an uncertainty which is a factor of two larger than
what enters the current PDG average.
More generally, several extractions exist which point to much lower values
of $\as(m_Z)$ than the PDG average. We mention in particular 
the extraction based on non-singlet structure functions from DIS \cite{Blumlein:2006be}, thrust \cite{Abbate:2010xh,Gehrmann:2012sc} and the C-parameter \cite{Hoang:2015hka} in $e^+e^-$ annihilation.
At the same time, PDF fits based on restricted data sets also tend
to lead to smaller values of $\as(m_Z)$. Whether this is a consequence or
specific theoretical assumptions or it points to a real issue, it is still a matter of debate.
Recently the PDF4LHC group has released the 2015 recommendation \cite{Butterworth:2015oua},
which, in particular, suggests to use, at $68\%$ CL
\begin{equation}
\as(m_Z)=0.1180\pm 0.0015\, .
\end{equation}
The slightly lower value with respect to the PDG determination, and, more importantly, the much larger uncertainty, should offer a more conservative standpoint for theoretical predictions in the next years.

\section{NLO and NLO+PS}
\label{sec:nlo}

Having discussed the PDFs entering the factorization theorem in Eq.~(\ref{ftheorem}), in the present and following Sections we come to discuss the computations of the partonic cross section.
In perturbative QCD, the cross section can be obtained as a power series expansion in $\as$
\begin{equation}
\label{sigmah}
d{\hat \sigma}_{ab}=\as^k\left(d{\hat \sigma}^{(0)}+\f{\as}{\pi}d{\hat \sigma}^{(1)}+\left(\f{\as}{\pi}\right)^2 d{\hat \sigma}^{(2)}+...\right)\, .
\end{equation}
Perturbative computations at the leading order (LO), i.e., limited to the first term in the expansion of Eq.~(\ref{sigmah}), are automated since quite some time \cite{Boos:2004kh,Alwall:2011uj,Mangano:2002ea,Gleisberg:2008fv,Cafarella:2007pc} but can only give the order of magnitude of cross section and kinematical distributions: the first meaningful result
is usually obtained at the next-to-leading order (NLO) in the QCD coupling $\as$. At NLO one has to consider real and virtual corrections that are separately infrared (IR) divergent: the singularities cancel when computing IR safe observables. Methods to achieve this cancellation at NLO are very well established
\cite{Frixione:1995ms,Frixione:1997np,Catani:1996jh,Catani:1996vz}.
For many years the bottleneck for NLO computations was the evaluation of
the relevant one-loop amplitudes. In recent years the traditional approach based on tensor integral reduction \cite{Passarino:1978jh,Denner:2002ii,Denner:2005nn} has been complemented with powerful methods based on recursion relations and unitarity \cite{Britto:2004ap,Ossola:2006us,Anastasiou:2006jv,Bern:2007dw,Ellis:2007br,vanHameren:2009vq}.
The high-energy community can now count on publicly available programs
such as {\sc Madgraph5\_aMC@NLO} \cite{Alwall:2014hca}, {\sc BlackHat} \cite{Bern:2012my}, {\sc GoSam} \cite{Cullen:2014yla}, {\sc Helac-NLO} \cite{Bevilacqua:2011xh},  {\sc NJet} \cite{Badger:2012pg}, {\sc Openloops} \cite{Cascioli:2011va}, {\sc Recola} \cite{Actis:2012qn}. These tools are sometimes combined with external codes for the subtraction of IR singularities like {\sc Helac-Dipoles} \cite{Czakon:2009ss}, {\sc MadFKS} \cite{Frederix:2009yq} and {\sc Sherpa} \cite{Gleisberg:2008ta}.
A nice example of the results one can obtain is shown in Table~\ref{tab:MG5} (from Ref.~\cite{Alwall:2014hca}), which reports LO and NLO cross sections for a sample from 172 processes obtained with {\sc Madgraph5\_aMC@NLO}.

%%%%%%%%%%%%%%%%%%%%%%%%%%%%%%%%%%%%%%%%%%%%%%%%%%%%%%%%%%%%%%%%%%%%%%%%
%%%%%%%%%%%%%%%%%%%%%%%%%%%%%%%%%%%%%%%%%%%%%%%%%%%%%%%%%%%%%%%%%%%%%%%%
\begin{table}%[p]
\begin{center}
\begin{tiny}
\begin{tabular}{l r@{$\,\to\,$}l lll}
\toprule
\multicolumn{3}{c}{Process~~~~~~~~~~~~~~~~~~~}& Syntax  & \multicolumn{2}{c}{Cross section (pb)}\\
\multicolumn{3}{c}{Single Higgs production~~~~~~~~~~~}& &  \multicolumn{1}{c}{  LO 13 TeV}&  \multicolumn{1}{c}{  NLO 13 TeV}\\
%%%%%%%%%%%%%%%%%%%%%%%%%%%%%%%%%%%%%%%%%%%%%%%%%%%%%%%%%%%%%%%%%%%%%%%%
\midrule
g.1 & $pp$ & $H$   (HEFT)         & {\tt p p > h } &  $ 1.593 \pm 0.003\, \cdot 10^{1} \quad {}^{+ 34.8 \% }_{- 26.0 \% } \,\, {}^{+  1.2 \% }_{-  1.7 \% }$ & $ 3.261 \pm 0.010\, \cdot 10^{1} \quad {}^{+ 20.2 \% }_{- 17.9 \% } \,\, {}^{+  1.1 \% }_{-  1.6 \% }$\\
g.2 & $pp$ & $Hj$      (HEFT)     & {\tt p p > h j } &  $ 8.367 \pm 0.003\, \cdot 10^{0} \quad {}^{+ 39.4 \% }_{- 26.4 \% } \,\, {}^{+  1.2 \% }_{-  1.4 \% }$ & $ 1.422 \pm 0.006\, \cdot 10^{1} \quad {}^{+ 18.5 \% }_{- 16.6 \% } \,\, {}^{+  1.1 \% }_{-  1.4 \% }$\\
g.3 & $pp$ & $Hjj$   (HEFT)       & {\tt p p > h j j} & $ 3.020 \pm 0.002\, \cdot 10^{0} \quad {}^{+ 59.1 \% }_{- 34.7 \% } \,\, {}^{+  1.4 \% }_{-  1.7 \% }$ & $ 5.124 \pm 0.020\, \cdot 10^{0} \quad {}^{+ 20.7 \% }_{- 21.0 \% } \,\, {}^{+  1.3 \% }_{-  1.5 \% } $\\
\midrule
g.4 & $pp$ & $Hjj$ (VBF)         & {\tt p p > h j j \$\$ w+ w- z} & $ 1.987 \pm 0.002\, \cdot 10^{0} \quad {}^{+  1.7 \% }_{-  2.0 \% } \,\, {}^{+  1.9 \% }_{-  1.4 \% }$ & $ 1.900 \pm 0.006\, \cdot 10^{0} \quad {}^{+  0.8 \% }_{-  0.9 \% } \,\, {}^{+  2.0 \% }_{-  1.5 \% }$ \\
g.5 & $pp$ & $Hjjj$ (VBF)        & {\tt p p > h j j j \$\$ w+ w- z} &  $ 2.824 \pm 0.005\, \cdot 10^{-1} \quad {}^{+ 15.7 \% }_{- 12.7 \% } \,\, {}^{+  1.5 \% }_{-  1.0 \% }$ & $ 3.085 \pm 0.010\, \cdot 10^{-1} \quad {}^{+  2.0 \% }_{-  3.0 \% } \,\, {}^{+  1.5 \% }_{-  1.1 \% }  $\\
\midrule
g.6 & $pp$ & $HW^\pm$         & {\tt p p > h wpm  } &  $ 1.195 \pm 0.002\, \cdot 10^{0} \quad {}^{+  3.5 \% }_{-  4.5 \% } \,\, {}^{+  1.9 \% }_{-  1.5 \% }$ & $ 1.419 \pm 0.005\, \cdot 10^{0} \quad {}^{+  2.1 \% }_{-  2.6 \% } \,\, {}^{+  1.9 \% }_{-  1.4 \% } $\\
g.7 & $pp$ & $HW^\pm\,j$      & {\tt p p > h wpm j } & $ 4.018 \pm 0.003\, \cdot 10^{-1} \quad {}^{+ 10.7 \% }_{-  9.3 \% } \,\, {}^{+  1.2 \% }_{-  0.9 \% }$ & $ 4.842 \pm 0.017\, \cdot 10^{-1} \quad {}^{+  3.6 \% }_{-  3.7 \% } \,\, {}^{+  1.2 \% }_{-  1.0 \% } $\\
g.8${}^*$ & $pp$ & $HW^\pm\,jj$     & {\tt p p > h wpm j j  } &  $ 1.198 \pm 0.016\, \cdot 10^{-1} \quad {}^{+ 26.1 \% }_{- 19.4 \% } \,\, {}^{+  0.8 \% }_{-  0.6 \% }$ & $ 1.574 \pm 0.014\, \cdot 10^{-1} \quad {}^{+  5.0 \% }_{-  6.5 \% } \,\, {}^{+  0.9 \% }_{-  0.6 \% }$\\
\midrule
g.9 & $pp$ & $HZ$       & {\tt p p > h z   } &  $ 6.468 \pm 0.008\, \cdot 10^{-1} \quad {}^{+  3.5 \% }_{-  4.5 \% } \,\, {}^{+  1.9 \% }_{-  1.4 \% }$ & $ 7.674 \pm 0.027\, \cdot 10^{-1} \quad {}^{+  2.0 \% }_{-  2.5 \% } \,\, {}^{+  1.9 \% }_{-  1.4 \% } $ \\
g.10 & $pp$ & $HZ\,j$    & {\tt p p > h z j  } &  $ 2.225 \pm 0.001\, \cdot 10^{-1} \quad {}^{+ 10.6 \% }_{-  9.2 \% } \,\, {}^{+  1.1 \% }_{-  0.8 \% }$ & $ 2.667 \pm 0.010\, \cdot 10^{-1} \quad {}^{+  3.5 \% }_{-  3.6 \% } \,\, {}^{+  1.1 \% }_{-  0.9 \% } $\\
g.11${}^*$ & $pp$ & $HZ\,j j$  & {\tt p p > h z j j  } & $ 7.262 \pm 0.012\, \cdot 10^{-2} \quad {}^{+ 26.2 \% }_{- 19.4 \% } \,\, {}^{+  0.7 \% }_{-  0.6 \% }$ & $  8.753 \pm 0.037\, \cdot 10^{-2} \quad {}^{+  4.8 \% }_{-  6.3 \% } \,\, {}^{+  0.7 \% }_{-  0.6 \% }  $\\
\midrule
g.12${}^*$ & $pp$ & $HW^+W^-$ (4f)   & {\tt p p > h w+ w- } &  $ 8.325 \pm 0.139\, \cdot 10^{-3} \quad {}^{+  0.0 \% }_{-  0.3 \% } \,\, {}^{+  2.0 \% }_{-  1.6 \% }$ & $ 1.065 \pm 0.003\, \cdot 10^{-2} \quad {}^{+  2.5 \% }_{-  1.9 \% } \,\, {}^{+  2.0 \% }_{-  1.5 \% }  $\\
g.13${}^*$ & $pp$ & $HW^\pm\gamma$    & {\tt p p > h wpm a } & $ 2.518 \pm 0.006\, \cdot 10^{-3} \quad {}^{+  0.7 \% }_{-  1.4 \% } \,\, {}^{+  1.9 \% }_{-  1.5 \% }$ & $ 3.309 \pm 0.011\, \cdot 10^{-3} \quad {}^{+  2.7 \% }_{-  2.0 \% } \,\, {}^{+  1.7 \% }_{-  1.4 \% }  $\\
g.14${}^*$ & $pp$ & $HZW^\pm$    & {\tt p p > h z wpm  } &  $ 3.763 \pm 0.007\, \cdot 10^{-3} \quad {}^{+  1.1 \% }_{-  1.5 \% } \,\, {}^{+  2.0 \% }_{-  1.6 \% }$ & $ 5.292 \pm 0.015\, \cdot 10^{-3} \quad {}^{+  3.9 \% }_{-  3.1 \% } \,\, {}^{+  1.8 \% }_{-  1.4 \% }  $\\
g.15${}^*$ & $pp$ & $HZZ$       & {\tt p p > h z  z } &  $ 2.093 \pm 0.003\, \cdot 10^{-3} \quad {}^{+  0.1 \% }_{-  0.6 \% } \,\, {}^{+  1.9 \% }_{-  1.5 \% }$ & $ 2.538 \pm 0.007\, \cdot 10^{-3} \quad {}^{+  1.9 \% }_{-  1.4 \% } \,\, {}^{+  2.0 \% }_{-  1.5 \% } $\\
\midrule
g.16 & $pp$ & $Ht\bar{t}$   & {\tt p p > h t t$\sim$  } &  $ 3.579 \pm 0.003\, \cdot 10^{-1} \quad {}^{+ 30.0 \% }_{- 21.5 \% } \,\, {}^{+  1.7 \% }_{-  2.0 \% }$ & $ 4.608 \pm 0.016\, \cdot 10^{-1} \quad {}^{+  5.7 \% }_{-  9.0 \% } \,\, {}^{+  2.0 \% }_{-  2.3 \% }  $\\
g.17 & $pp$ & $Htj $    & {\tt p p > h tt j   } &  $ 4.994 \pm 0.005\, \cdot 10^{-2} \quad {}^{+  2.4 \% }_{-  4.2 \% } \,\, {}^{+  1.2 \% }_{-  1.3 \% }$ & $ 6.328 \pm 0.022\, \cdot 10^{-2} \quad {}^{+  2.9 \% }_{-  1.8 \% } \,\, {}^{+  1.5 \% }_{-  1.6 \% } $\\
g.18 & $pp$ & $Hb\bar{b}$ (4f) & {\tt p p > h b b$\sim$} &$ 4.983 \pm 0.002\, \cdot 10^{-1} \quad {}^{+ 28.1 \% }_{- 21.0 \% } \,\, {}^{+  1.5 \% }_{-  1.8 \% }$& $ 6.085 \pm 0.026\, \cdot 10^{-1} \quad {}^{+  7.3 \% }_{-  9.6 \% } \,\, {}^{+  1.6 \% }_{-  2.0 \% }$\\
\midrule
g.19 & $pp$ & $Ht\bar{t}j$   & {\tt p p > h t t$\sim$ j } &$ 2.674 \pm 0.041\, \cdot 10^{-1} \quad {}^{+ 45.6 \% }_{- 29.2 \% } \,\, {}^{+  2.6 \% }_{-  2.9 \% }$& $ 3.244 \pm 0.025\, \cdot 10^{-1} \quad {}^{+  3.5 \% }_{-  8.7 \% } \,\, {}^{+  2.5 \% }_{-  2.9 \% }  $\\
g.20${}^*$ & $pp$ & $Hb\bar{b}j$ (4f) & {\tt p p > h b b$\sim$ j } &$ 7.367 \pm 0.002\, \cdot 10^{-2} \quad {}^{+ 45.6 \% }_{- 29.1 \% } \,\, {}^{+  1.8 \% }_{-  2.1 \% }$& $ 9.034 \pm 0.032\, \cdot 10^{-2} \quad {}^{+  7.9 \% }_{- 11.0 \% } \,\, {}^{+  1.8 \% }_{-  2.2 \% }$\\
\bottomrule
%%%%%%%%%%%%%%%%%%%%%%%%%%%%%%%%%%%%%%%%%%%%%%%%%%%%%%%%%%%%%%%%%%%%%%%%
\end{tabular}
\end{tiny}
\end{center}
\caption{\label{tab:MG5} 
Sample of LO and NLO predictions for the production of a single SM Higgs,
possibly in association with other SM particles and with cuts, at the LHC with $\sqrt{s}=13$ TeV. Scale and PDFs uncertainties are also reported (from Ref.~\cite{Alwall:2014hca}).
}
\end{table}
%\end{sidewaystable}
%%%%%%%%%%%%%%%%%%%%%%%%%%%%%%%%%%%%%%%%%%%%%%%%%%%%%%%%%%%%%%%%%%%%%%%%

We can thus conclude that the problem of doing NLO calculations is {\em in principle} solved: what remains to be done is an extensive validation of the available codes, by comparing predictions from the different tools and assessing their limitations
in terms of final state complexity. Further theoretical improvements in terms of
performance will be important to allow us to fully exploit the developments discussed above.

Even including higher-order corrections, parton level calculations
cannot provide a realistic description of actual events at hadron colliders.
Monte Carlo (MC) event generators are indispensable tools for data analysis
at high-energy collider experiments. Through the complete simulation of all the stages of the hadronic collision, from the initial state radiation to the hard scattering process (including secondary scattering processes, the so called underlying event) to the final state radiation, the hadronization and the detector resolution effects, they provide a realistic description of the hadronic collisions. In the past, MC generators were based on a set of underlying processes, computed at LO, on top of which the initial and final state QCD radiation was built up, by using the Monte Carlo parton shower (PS): a cascade of subsequent multiparton emissions is generated using soft and collinear approximations. Being essentially LO, such generators could hardly describe an event with one or more additional (w.r.t. the LO) hard jets, since QCD radiation was obtained through the parton shower.
In the last decades, MC event generators have benefitted from two major theoretical inprovements.
The first one is the possibility to merge QCD multiparton matrix elements with the parton shower (ME+PS),
so as to correctly account for the emission of further hard partons \cite{Catani:2001cc}.
This approach has been developed and implemented
in several variants (for a review, see e.g. Ref.~\cite{Buckley:2011ms}).
The second is the possibility to consistently include exact NLO corrections in the simulation,
so as to achieve a NLO accurate generator (NLO+PS).
Two approaches exist to carry out this procedure: the {\sc MC@NLO} \cite{Frixione:2002ik} and {\sc POWHEG} \cite{Frixione:2007vw} methods, which, roughly speaking, differ in the amount of radiation which is actually exponentiated. Although the two approaches lead to simulations with the same formal accuracy,
there are quantitative differences which may be significant. A current issue is indeed the one
of quantifying the uncertainties in NLO+PS simulations.
State of the art NLO+PS simulations with the MC@NLO method can be obtained with {\sc MadGraph5\_aMC@NLO},
{\sc Openloops+SHERPA} \cite{Cascioli:2011va,Hoeche:2011fd}, {\sc Herwig++Matchbox} \cite{Platzer:2011bc}.
The POWHEG method is implemented in the
{\sc POWHEGBox} \cite{Alioli:2010xd} and in {\sc Herwig++Matchbox} \cite{Platzer:2011bc}.

Starting from NLO+PS matching, the next step to improve MC generators is either adding higher-multiplicity tree-level matrix elements \cite{Hamilton:2010wh}, or merging NLO+PS simulations with different multiplicities \cite{Hoeche:2012yf,Frederix:2012ps,Platzer:2012bs,Lonnblad:2012ix}.
An interesting possibility is the one offered by the {\sc Minlo} \cite{Hamilton:2012np} and {\sc Geneva} \cite{Alioli:2012fc} approaches, which avoid the use of a merging scale, and pave the way to NNLO+PS simulations, to be discussed in Sect.~\ref{sec:nnlo}.

A proposal for an alternative NLO+PS scheme, which, however, requires dedicated PDFs is described in \cite{Jadach:2015mza}.
Other improvements \cite{Platzer:2013fha,Nagy:2014mqa} try to overcome the traditional limitations of MC generators. The showers in fact usually employ a large-$N_c$ approximation to obtain a probabilistic description and to be able to neglect quantum interferences.

\section{Resummation}

In processes where multiple (and different) scales are present fixed order perturbative computations develop large logarithmic terms of IR origin that can
spoil the perturbative expansion. To obtain reliable predictions these terms must be resummed to all orders. 
The parton shower event generators discussed in the previous section
carry out this resummation to a limited (essentially leading) logarithmic accuracy.
For specific observables, analytic resummation reaches
a higher logarithmic accuracy, and thus provides better theoretical predictions,
that can also be used to validate MC tools.
A typical example is the transverse momentum spectrum of the Higgs boson,
for which resummed predictions at
next-to-next-to-leading logarithmic (NNLL) accuracy \cite{Bozzi:2005wk,deFlorian:2011xf}, matched to the NNLO result valid at large $p_T$ are normally used to correct (reweight) results from MC event generators.
The calculation of Ref.~\cite{deFlorian:2012mx,Grazzini:2013mca}, implemented in the numerical program {\sc HRes}, includes the possibility to study the relevant Higgs boson decays.
An analogous computation for vector boson production has been recently implemented in the
numerical program {\sc DYres} \cite{Catani:2015vma}, and it allows us to study vector boson production
including the leptonic decay.

\begin{figure}[htpb]
    \begin{subfigure}[b]{0.42\textwidth}
        \centering
        \includegraphics[width=\textwidth]{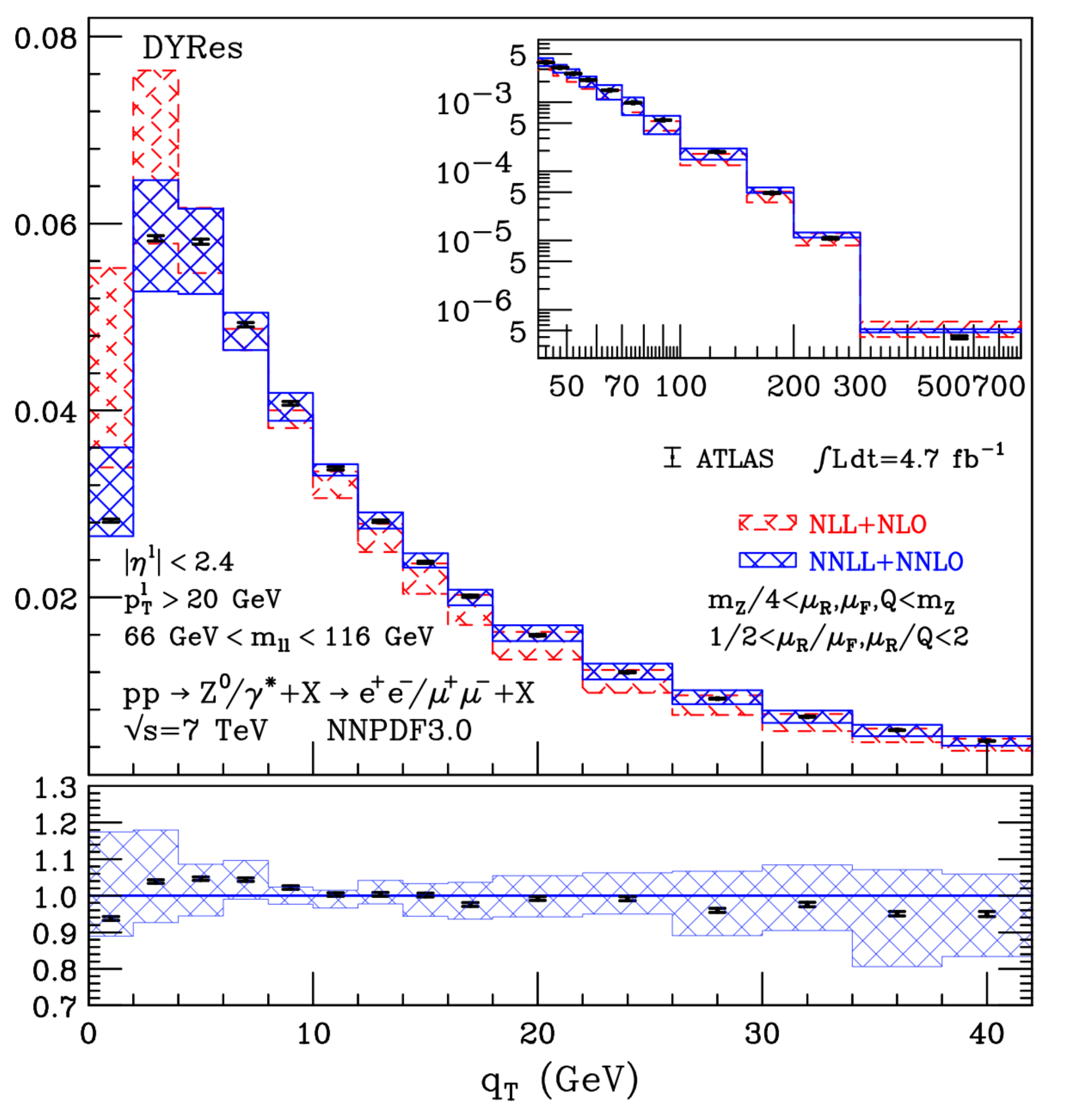}
    \end{subfigure}
    \begin{subfigure}[b]{0.56\textwidth}
        \centering
        \includegraphics[width=\textwidth]{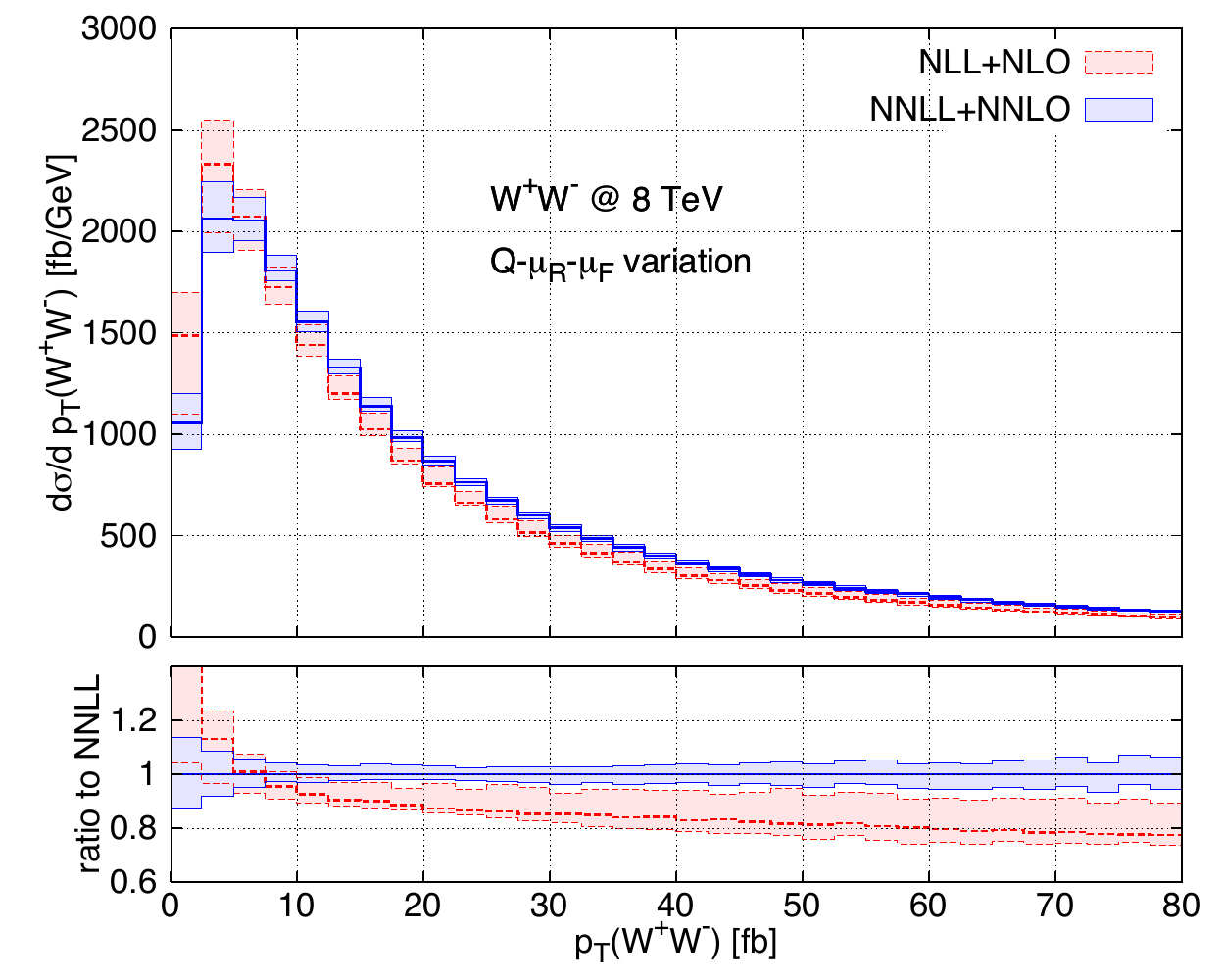}
    \end{subfigure}
    \caption{Transverse momentum spectrum of a $Z$ boson compared to ATLAS data (left) (from Ref.~\cite{Catani:2015vma}) of a $WW$ pair (right) (from Ref.~\cite{Grazzini:2015wpa}).
}
    \label{fig:qtresum}
\end{figure}

In Fig.~\ref{fig:qtresum} (left) a comparison of the NNLL+NNLO result obtained with {\sc DYRes} to ATLAS data
is shown (from Ref.~\cite{Catani:2015vma}). The calculation is carried out in the same fiducial region
in which the measurement is performed, and it nicely describes the data, within perturbative uncertainties.
Based on the universality properties of soft and collinear emissions, and
on the complete understanding of the structure of the resummation coefficients for arbitrary
colour singlet processes \cite{Catani:2013tia}, transverse momentum resummation can be generalized to
other processes once the relevant amplitudes are known.
In Fig.~\ref{fig:qtresum} (right) (from Ref.~\cite{Grazzini:2015wpa})
predictions for the transverse momentum spectrum of $WW$ pairs
are presented, for the first time, at full NNLL+NNLO accuracy.

An important class of observables for which resummation has been studied
for long time are event shapes in $e^+e^-$ collisions,
for which the resummation has been carried out on an observable dependent basis
(see e.g. Ref.~\cite{Luisoni:2015xha} and references therein).
Automated (observable-independent) resummation for $e^+e^-$ event shape variables
fulfilling a specific (more restrictive) criterion of IR safety, was first addressed up to NLL accuracy
in Ref.~\cite{Banfi:2001bz}, by using a seminumerical method.
Recently, this method has been extended to NNLL
in Ref.~\cite{Banfi:2014sua}. The approach used for these calculations is potentially viable also for hadron collisions.

%In recent years the traditional approach to QCD resummations, based on soft and collinear factorization, has been complemented with a new approach based on Soft Collinear Effective Theory (SCET) \cite{Bauer:2000ew,Bauer:2000yr,Bauer:2001ct,Bauer:2001yt}.

In recent years the traditional approach to QCD resummations has been complemented with a new approach
based on Soft Collinear Effective Theory (SCET) \cite{Bauer:2000ew,Bauer:2000yr,Bauer:2001ct,Bauer:2001yt}. Traditional resummation uses soft and collinear factorization formulae for scattering amplitudes \cite{Bassetto:1984ik,Kosower:1999xi,Bern:1999ry,Catani:1999ss,Catani:2000pi}, while SCET uses non-local effective operators to describe the relevant IR modes.
A recent application of SCET resummation is the resummed computation of the $C$ parameter \cite{Hoang:2015hka}, which leads to an extraction of $\as(m_Z)=0.1123\pm 0.0015$,
consistent with a previous determination based on thrust \cite{Abbate:2010xh}\footnote{This is not surprising, since these two event shape variables are strongly correlated.}.
Other important results regard the so called $N-$jettiness variable $\tau_N$ \cite{Stewart:2010tn}. $N$-jettiness is an event shape variable in hadron collisions
which vanishes in the limit in which there are exactly $N$ narrow jets.
In the limit $\tau_N\ll 1$ additional QCD radiation is strongly constrained and the differential cross section can be written as
\begin{equation}
\sigma(\tau_N<\tau_{\rm cut})\sim\int H\otimes B_1\otimes B_2\otimes \prod_{i=1}^N J_n\otimes S\, .
\end{equation}
The beam functions $B_1$ and $B_2$ embody collinear radiation to the initial state hadrons. The jet functions embody
collinear radiation in the direction of the outgoing jets, the hard function contain
virtual effects, and the soft function describes the additional soft radiation.

The jet functions are known up to ${\cal O}(\as^2)$ \cite{Becher:2006qw,Becher:2010pd}.
The beam functions are also known at the same order \cite{Gaunt:2014xga,Gaunt:2014cfa}.
In the case of $\tau_1$ the soft function has been computed up to relative order ${\cal O}(\as^2)$ \cite{Boughezal:2015eha}, and
thus the resummation can be carried out at full NNLL accuracy, having control of all the terms of relative order ${\cal O}(\as^2)$ at small $\tau_1$. An interesting spin-off of these results regards the application to fully exclusive NNLO calculations, that will be discussed in the next section.

\section{NNLO and beyond}
\label{sec:nnlo}

In Sec.~\ref{sec:nlo} we have stressed that NLO is the first order at which a perturbative
QCD result can be considered reliable. This implies that it is with NNLO that we
obtain a first reliable estimate of the uncertainty. More generally, the inclusion
of NNLO corrections is important {\em at least} in the following cases: i) benchmark processes measured with high accuracy. At hadron colliders these include $pp\to W(Z)$ production, $pp\to t{\bar t}$, $pp\to {\rm jet(s)}$; ii) processes for which NLO corrections are particularly large. These include single Higgs production, possibly accompanyed by one jet, double Higgs production; iii) important backgrounds for Higgs and new physics searches. These include, besides the aforementioned processes, all the vector boson pair production processes, plus, for example, $W(Z)+{\rm jet(s)}$.

While general techniques for NLO calculations have been available for long time,
the extension of perturbative QCD calculations to the NNLO
is definitely a complicated task. First of all, one needs the relevant QCD amplitudes.
While the computation of tree-level and one loop amplitudes is nowadays essentially automated, the computation of two-loop amplitudes is at the frontier of current techniques.
The massless QCD amplitudes for all the $2\to 2$ parton processes are known since long time \cite{Anastasiou:2000kg,Anastasiou:2000ue,Anastasiou:2001sv,Glover:2001af,Bern:2002tk}. Recently, the computation of the helicity amplitudes for all the diboson production processes was completed \cite{Gehrmann:2011ab,Caola:2014iua,Gehrmann:2015ora}. The two-loop amplitudes for Higgs \cite{Gehrmann:2011aa} and vector \cite{Garland:2001tf} boson production in association with a jet are also available. On the contrary, the two-loop amplitudes
for heavy-quark production are available only in numerical form \cite{Czakon:2008zk,Baernreuther:2013caa}, while analytical results exist for some of the colour factors \cite{Bonciani:2008az,Bonciani:2009nb,Bonciani:2010mn,Bonciani:2013ywa}.

Even having all the relevant amplitudes, the cancellation of the IR singularities and the implementation into a numerical program able to compute cross sections and distributions is still a highly non-trivial task.
Various methods have been proposed to achieve IR cancellations and to set up a NNLO calculation.
These methods can be divided into two broad categories.
In the first one, one organizes the NNLO calculation from scratch,
trying to cancel IR singularities of {\em both} NLO and NNLO type. 
The formalisms of {\em antenna} subtraction 
\cite{Kosower:1997zr, GehrmannDeRidder:2005cm,Daleo:2006xa,Currie:2013vh},
{\em colourful} subtraction \cite{Somogyi:2005xz, DelDuca:2015zqa}
and {\em Stripper} \cite{Czakon:2010td,Czakon:2011ve,Czakon:2014oma}
belong to this category. Antenna subtraction and colorful subtraction
can be considered as extensions of the subtraction method \cite{Frixione:1995ms,Frixione:1997np,Catani:1996jh,Catani:1996vz} at NNLO.
Stripper is instead a combination of the subtraction method with numerical techniques based on 
{\em sector decomposition} \cite{Anastasiou:2003gr,Binoth:2000ps}.
The methods in the second category start from a NLO calculation with higher multiplicity
and device suitable subtractions to make the cross section
finite in the region in which the additional parton (jet) leads
to further divergences.
Typical examples in this class are the
$q_T$ subtraction formalism \cite{Catani:2007vq} and the recently proposed 
$N$-{\em jettiness} subtraction 
\cite{Boughezal:2015dva,Boughezal:2015eha,Gaunt:2015pea}. The above methods
exploit the knowledge of the singular behavior of the cross section at small $q_T$ (${\tau_N}$)
to set up appropriate
counterterms to cancel the IR singularities.
This knowledge is obtained from the resummations discussed in the previous section.
Another method belonging to this class is the recently proposed {\em Born projection} method,
used to perform the NNLO computation for vector boson fusion (VBF) \cite{Cacciari:2015jma}.
In the last few years we have witnessed an impressive series of new NNLO calculations:
in the following we briefly review some of the most interesting results.

\begin{figure}[htpb]
    \begin{subfigure}[b]{0.50\textwidth}
        \centering
        \includegraphics[width=\textwidth]{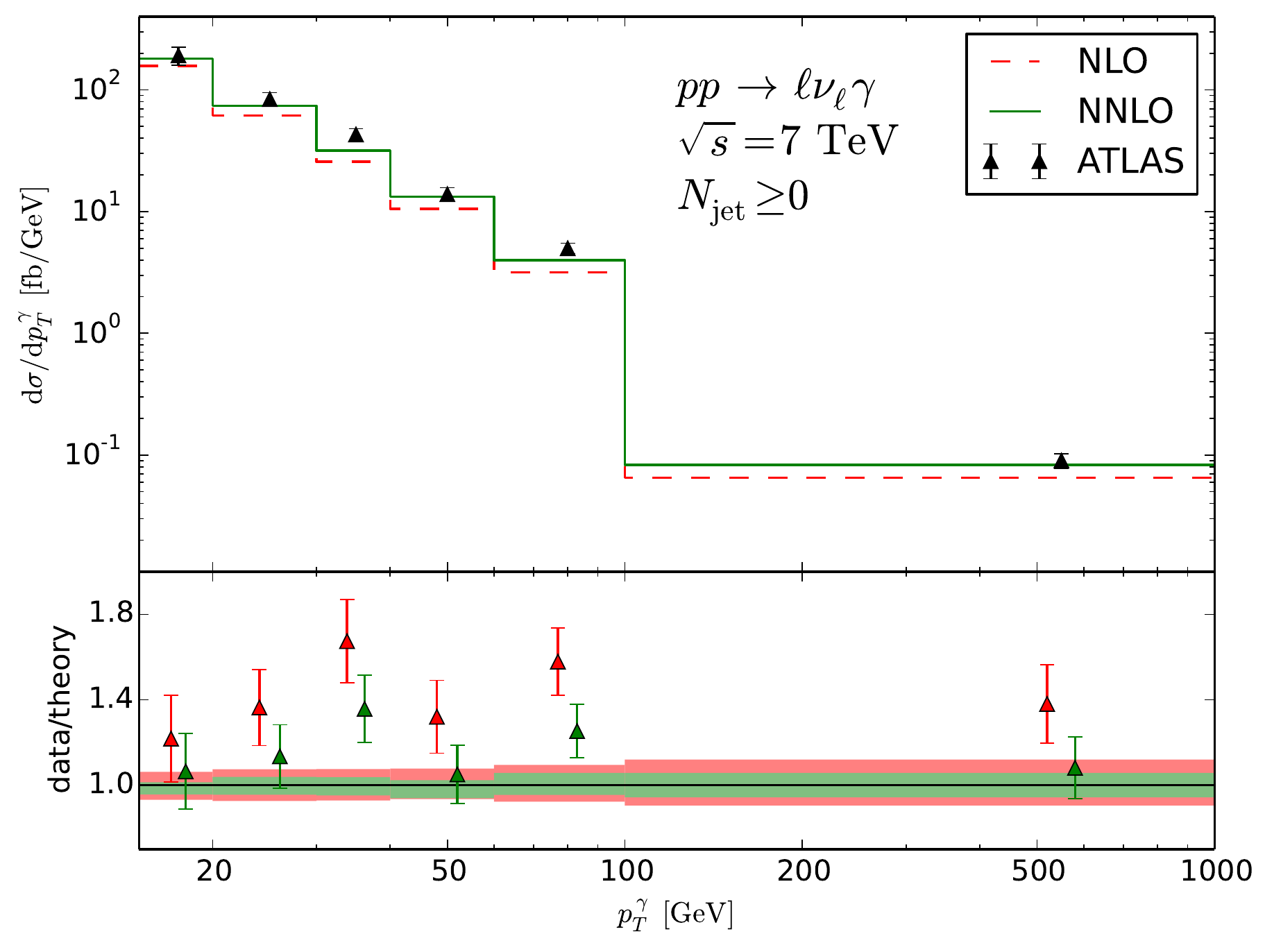}
    \end{subfigure}
    \begin{subfigure}[b]{0.48\textwidth}
        \centering
        \includegraphics[width=\textwidth]{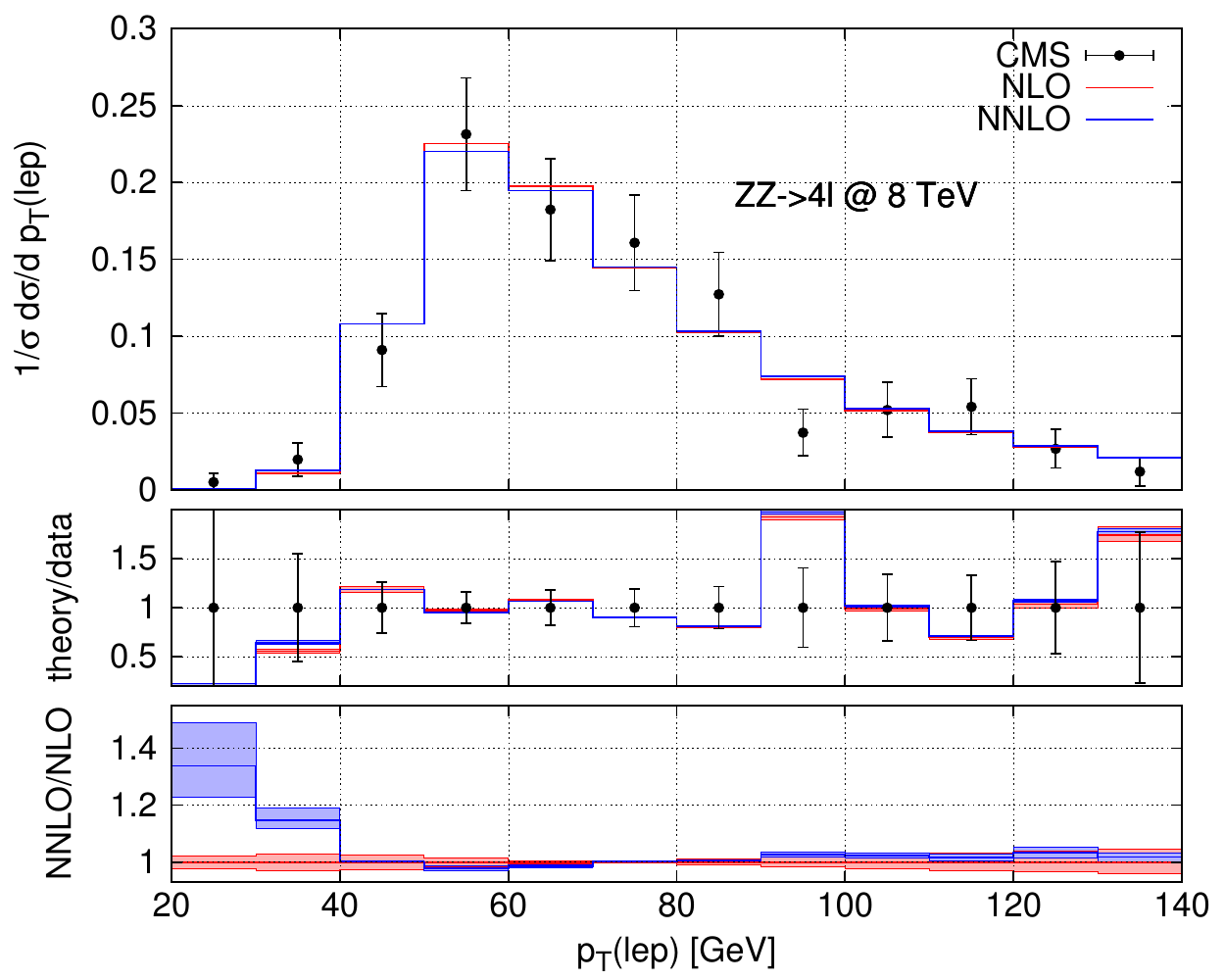}
    \end{subfigure}
    \caption{
The photon $p_T$ distribution in $l\nu_l\gamma$ production compared to ATLAS data (left) (from Ref.~\cite{Grazzini:2015nwa}).
The $p_T$ distribution of the leading lepton in four lepton production at 8 TeV at NLO and NNLO compared to the CMS data (right) (from Ref.~\cite{Grazzini:2015hta}).    
}
    \label{fig:zzwg}
\end{figure}

The previously mentioned computation of the two-loop amplitudes for vector boson pair production allowed us to use the $q_T$ subtraction method to carry out various
NNLO computations \cite{Cascioli:2014yka,Gehrmann:2014fva,Grazzini:2015nwa,Grazzini:2015hta}.
As an example of these results, in Fig.~\ref{fig:zzwg} (left) we report result of
the NNLO calculation the photon $p_T$ distribution in $W\gamma$ production at NLO and NNLO compared
to ATLAS data (from Ref.~\cite{Grazzini:2015nwa}). The NNLO corrections are quite large, being at the ${\cal O}(20\%)$ level, and significantly improve the agreement with the data.

In the case of four-lepton production, in Fig.~\ref{fig:zzwg} (right)
we report result of the NNLO calculation
of the leading lepton $p_T$ distribution in four lepton production (from Ref.~\cite{Grazzini:2015hta})
compared to the avaliable CMS data. The theoretical prediction agrees with the data, which, however, have still large uncertainties. We stress that both
calculations are carried out in the fiducial region where
the measurements are performed.

\begin{figure*}
  \centering
          \includegraphics[clip,height=0.45\textwidth,page=1,angle=0]{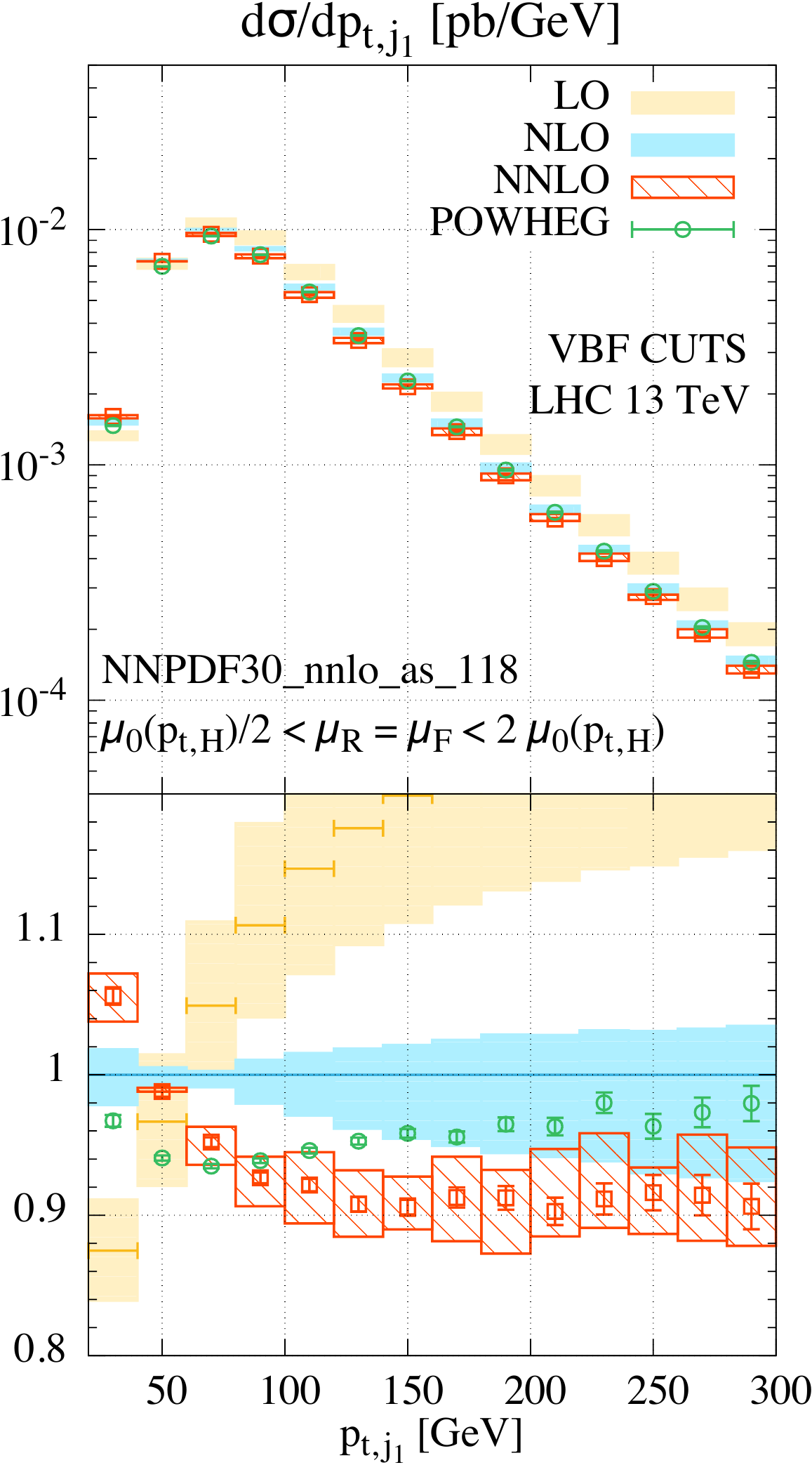}%
    \hfill\includegraphics[clip,height=0.45\textwidth,page=2,angle=0]{NLO-NNLO-crop.pdf}%
    \hfill\includegraphics[clip,height=0.45\textwidth,page=3,angle=0]{NLO-NNLO-crop.pdf}\hspace{0.8mm}%
    \hfill\includegraphics[clip,height=0.45\textwidth,page=4,angle=0]{NLO-NNLO-crop.pdf}%
  \caption{From left to right, differential cross sections for the transverse momentum
distributions  for the two leading jets,
$p_{t,j_1}$ and $p_{t,j_2}$, for the Higgs boson, $p_{t,H}$, and the distribution for the rapidity 
separation between the two leading jets, $\Delta y_{j_1,j_2}$ (from Ref.~\cite{Cacciari:2015jma}). }
  \label{fig:vbf}
\end{figure*}

The NNLO computation of VBF leads to particularly interesting results \cite{Cacciari:2015jma}. NNLO corrections in the structure function approach (neglecting color connections between the two quark lines) were first computed at fully inclusive level, and found to be very small \cite{Bolzoni:2010xr}.
The fully exclusive computation of Ref.~\cite{Cacciari:2015jma} shows that the NNLO effect
is larger when VBF cuts are applied and can be at the ${\cal O}(5\%)$ level (see Fig.~\ref{fig:vbf}). This is indeed the reason why using $K$-factors
obtained from inclusive NNLO calculations may lead to a mismodelling of radiative effects.

%----------------------------------------------------------------------
%----------------------------------------------------------------------

Another important ongoing NNLO calculation is the one of jet production. As discussed in Sec.~\ref{sec:pdfs}, NNLO predictions for this process will allow a consistent inclusion of jet data in NNLO PDF fits.
After the completion
of the $gg$ contribution  \cite{Currie:2013dwa}, first results for the $qg$ and $q{\bar q}$ channels in the leading color approximation have been presented recently \cite{currie}.

As far as top production is concerned, the computation of the total cross section \cite{Czakon:2013goa}
has been recently followed by the calculation of the $t{\bar t}$ asymmetry at the Tevatron \cite{Czakon:2014xsa}, and, very recently,
other differential results appeared \cite{Czakon:2015owf}.
The above results are obtained with the Stripper method \cite{Czakon:2010td,Czakon:2011ve,Czakon:2014oma}.

Other recent important results regard Higgs and vector boson in association with a jet.
The computation of $W+{\rm jet}$ at NNLO has been completed this year using $N$-jettiness \cite{Boughezal:2015dva}.
The NNLO corrections are small and lead to a significant reduction of perturbative uncertainties.
Parallely, the NNLO computation of $Z+{\rm jet}$ has been completed by using the antenna subtraction method \cite{Ridder:2015dxa}. Such calculation, currently limited to the leading color approximation, also finds moderate corrections, and will be also important to improve the theoretical predictions for the $p_T$ spectrum of the $Z$ boson.
The NNLO calculation for $H+{\rm jet}$ has been carried out with three independent methods \cite{Chen:2014gva,Boughezal:2015dra,Boughezal:2015aha}.

\begin{figure}[htpb]
    \begin{subfigure}[b]{0.53\textwidth}
        \centering
        \includegraphics[width=\textwidth]{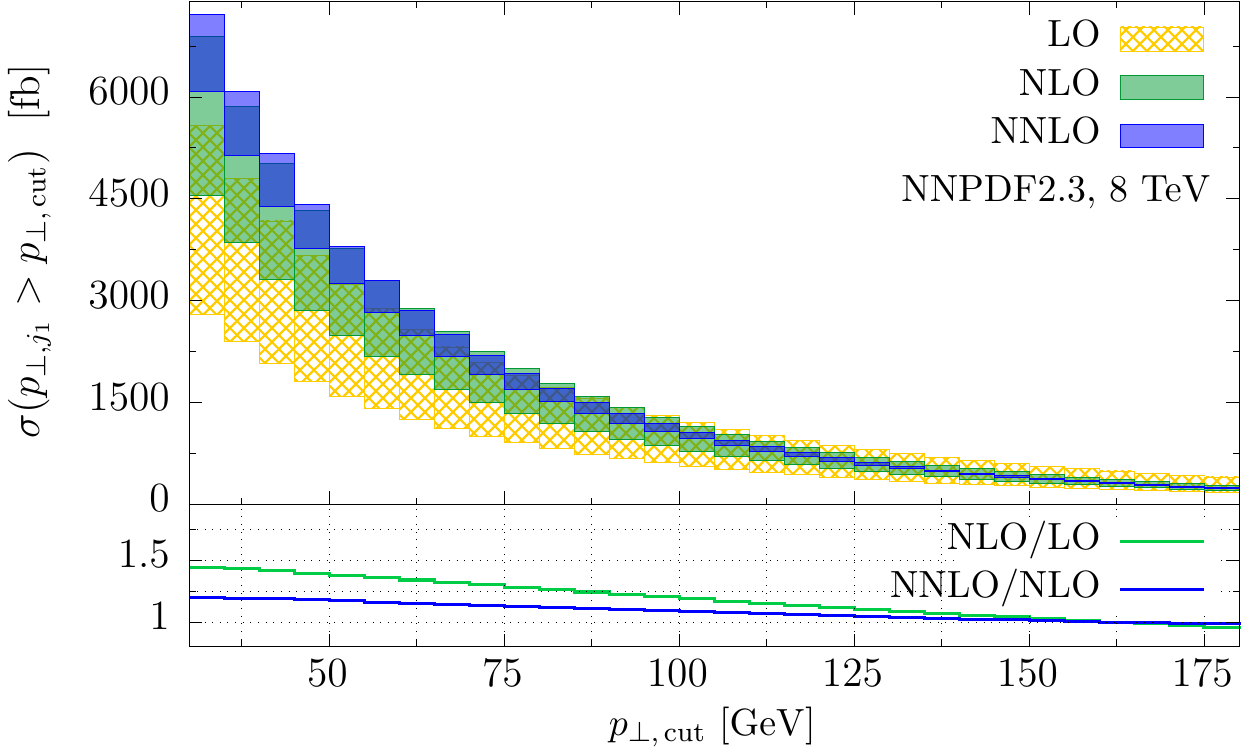}
    \end{subfigure}
    \begin{subfigure}[b]{0.42\textwidth}
        \centering
        \includegraphics[width=\textwidth]{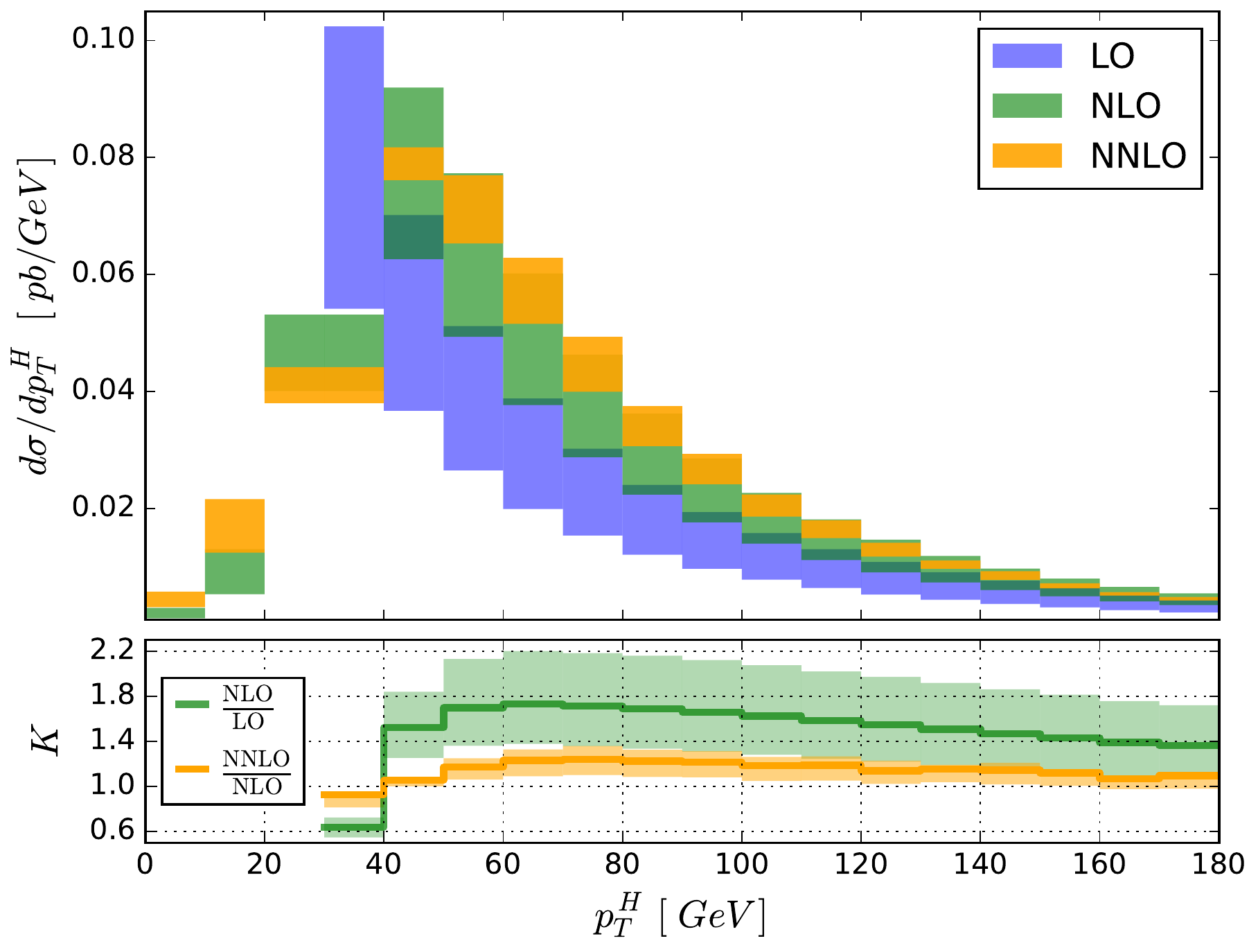}
    \end{subfigure}
    \caption{
The cross section with $p_T^{\rm jet}>p_T^{\rm cut}$ (left) (from Ref.~\cite{Boughezal:2015dra}) and
the Higgs $p_T$ spectrum at NNLO (right) (from Ref.~\cite{Boughezal:2015aha}).
}
    \label{fig:hjet}
\end{figure}

Higgs production at high-$p_T$ is useful to test new physics scenarios and better resolve the structure of the heavy-quark loop.
The NNLO calculation is performed in the large-$m_{top}$ approximation, and this is the approximation in which also NLO corrections are currently available.
The inclusion of finite top-mass effects would be important, but requires two-loop computations with different mass scales, which are at the boundary of current techniques.
As far as the inclusive cross section for Higgs boson production in gluon fusion is concerned
even the N$^3$LO corrections have been recently computed \cite{Anastasiou:2015ema} in the
large-$m_{top}$ approximation. This computation, which represents the first result at this perturbative order in hadronic collisions, leads to a small corrections with respect to NNLO but to a significant reduction of perturbative uncertainties.

The N$^3$LO result for the inclusive cross section and the NNLO calculation of the $H+{\rm jet}$ cross section
can be combined to obtain the corresponding cross section with a jet veto \cite{Catani:2001cr}. This step has been recently carried out
in Ref.~\cite{Banfi:2015pju}, where the effect of NNLL resummation of the jet $p_T$ and the LL resummation of the
jet radius are also included. The results of Ref.~\cite{Banfi:2015pju} set a new reference for theoretical predictions for the jet vetoed cross section and the corresponding efficiency, with perturbative uncertainties which are down to the few percent level.

With the impressive amount of new NNLO results that appeared in this year, and, at the same time, the
existence of established schemes to match NLO computations to PS simulations, a natural question
is whether one can match NNLO computations to PS simulations.
This topic has received considerable attention recently, and three methods have been proposed to deal with the production of a colorless high-mass system $F$ (e.g. vector and Higgs boson production).
The first method \cite{Hamilton:2012rf} uses the {\sc Minlo} approach \cite{Hamilton:2012np} to merge two inclusive NLO+PS samples for $F$ and $F+{\rm jet}$ production, and then uses the NNLO information on the rapidity distribution of $F$ obtained from NNLO parton level generators \cite{Catani:2007vq,Grazzini:2008tf,Catani:2009sm} to enforce
the correct NNLO normalization through a reweighting procedure. The method has been applied to Higgs \cite{Hamilton:2013fea} and vector boson production \cite{Karlberg:2014qua}.
The second method, based on an improved version of the merging scheme of Ref.~\cite{Lonnblad:2012ix} and
on a variant of the NNLO subtraction method of Ref.~\cite{Catani:2007vq}, has also been applied to vector \cite{Hoeche:2014aia} and Higgs boson production \cite{Hoche:2014dla}.
A third method, based on the {\sc GENEVA} merging approach \cite{Alioli:2012fc}, has been recently applied to vector boson production \cite{Alioli:2015toa}. It will be interesting to see how the results obtained with the three different approaches compare to each other.

\section{Electroweak corrections}

We finally turn to discuss Electroweak (EW) corrections.
A naive power counting argument suggests that, since ${\cal O}(\alpha)\sim {\cal O}(\as^2)$,
EW corrections should be of the same order as NNLO QCD corrections.
This means that, when precision is an issue, the inclusion of EW corrections is mandatory. A typical example
is the $m_W$ measurement planned at the LHC. To achieve a target accuracy of $\Delta m_W\sim 10$ MeV,
extremely accurate predictions are needed for the relevant kinematical distributions like the lepton $p_T$ and the transverse mass. NLO EW corrections \cite{Zykunov:2001mn,Dittmaier:2001ay,Baur:2004ig} lead to important effects and are often included by
assuming complete factorization (see e.g. Ref.~\cite{Li:2012wna}).
Work towards mixed ${\cal O}(\alpha\alpha_S)$ corrections is ongoing \cite{Dittmaier:2014qza,Dittmaier:2015rxo}.

\begin{figure}[htpb]
    \begin{subfigure}[b]{0.48\textwidth}
        \centering
        \includegraphics[width=\textwidth]{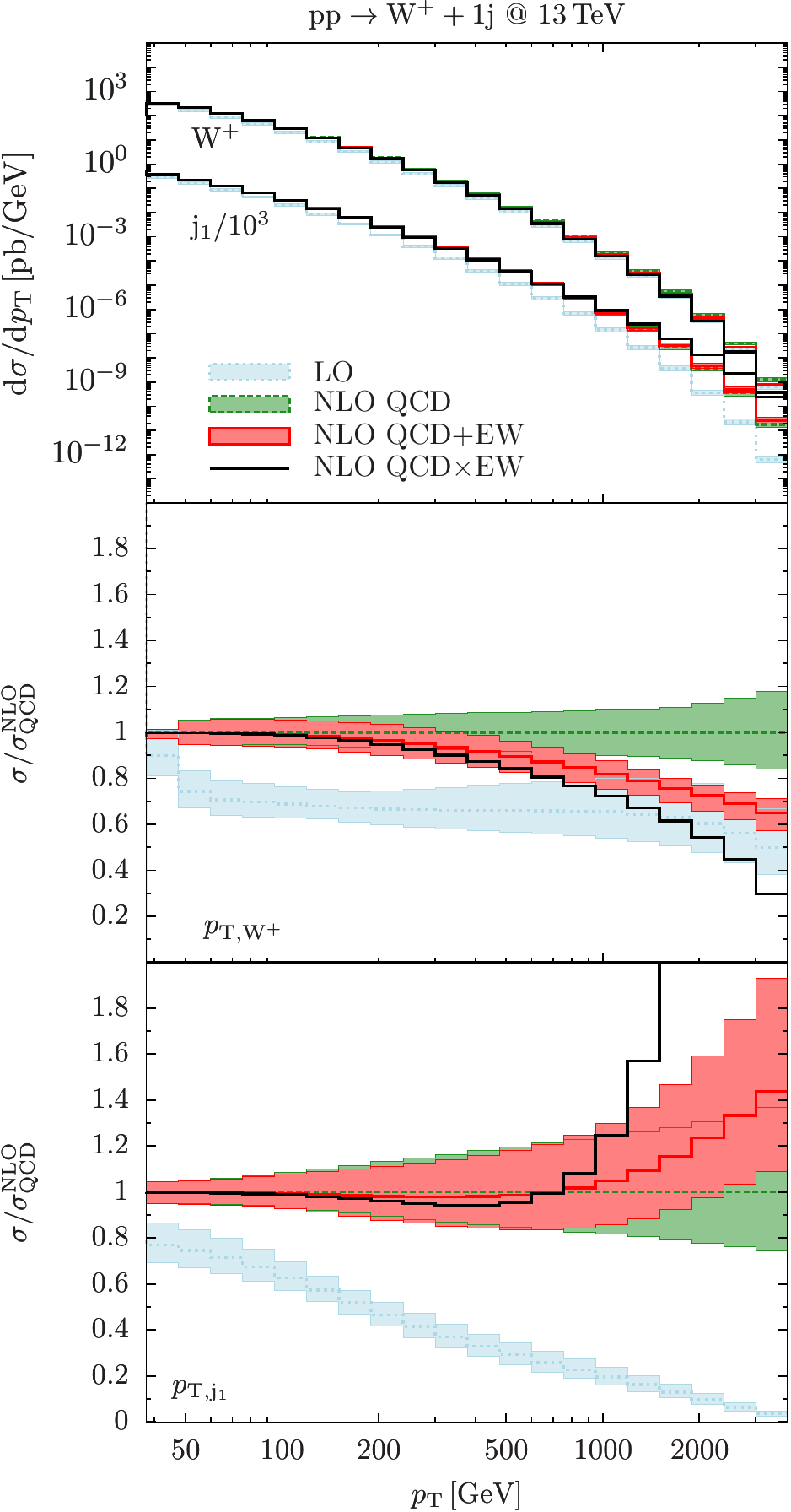}
    \end{subfigure}
    \begin{subfigure}[b]{0.48\textwidth}
        \centering
        \includegraphics[width=\textwidth]{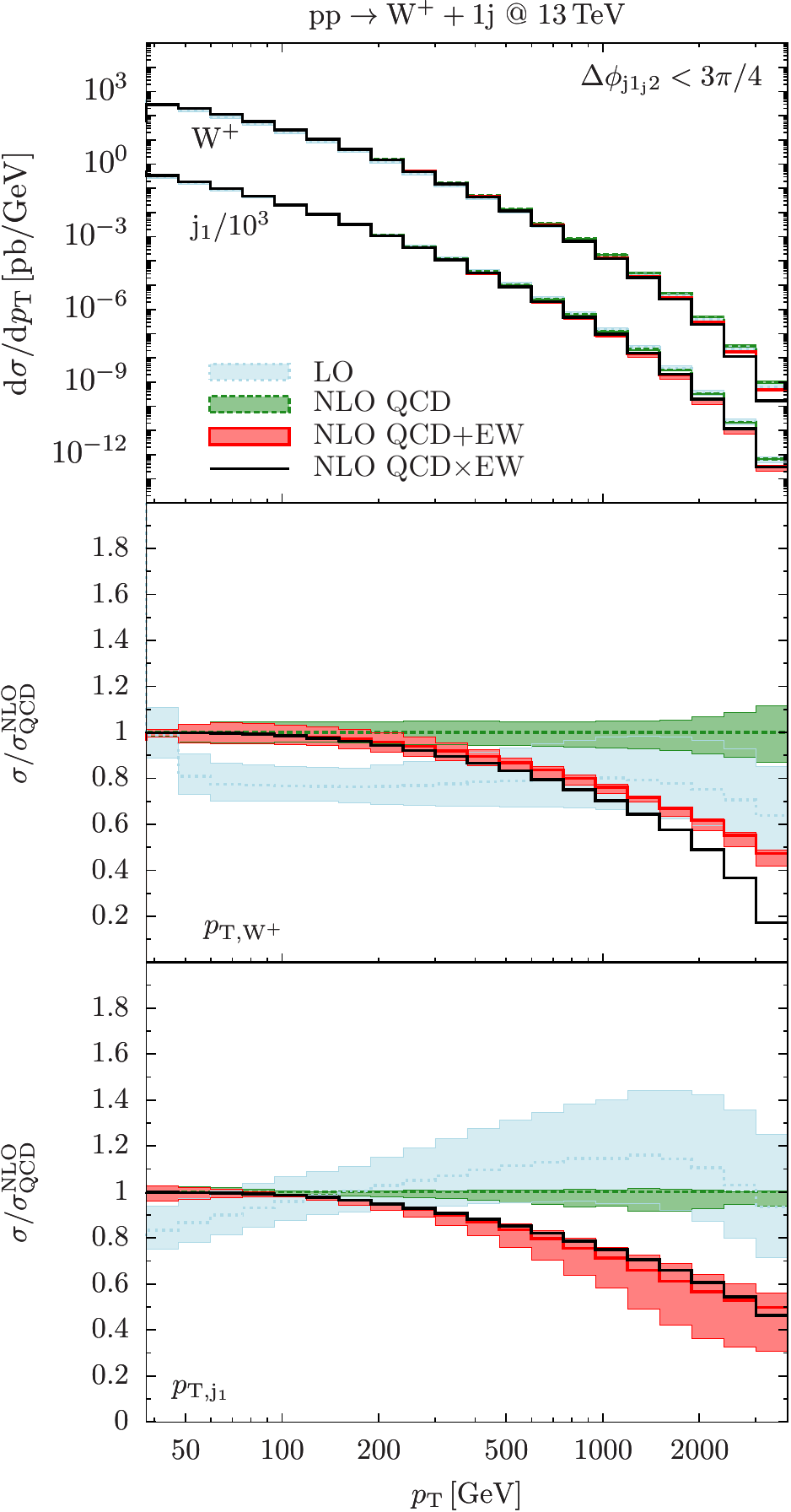}
    \end{subfigure}
    \caption{Transverse momentum spectra of the $W$ boson and of the leading jet for inclusive $W^++1$ jet production (left) and with a cut on the azimuthal separation of the first two jets (from Ref.~\cite{Kallweit:2014xda}).
}
    \label{fig:ew}
\end{figure}

Besides the impact in precision measurements,
it is well known \cite{Ciafaloni:1998xg,Denner:2000jv} that EW effects become increasingly important at high energies.
In this regime large logarithmic contributions of Sudakov \cite{Sudakov:1954sw} type appear
involving the ratio of the energy scale over the electroweak scale. In massless gauge theories like QCD the singularities in the virtual corrections have to be cancelled by adding the real emission contributions.
In the EW theory the gauge boson masses provide a physical cutoff, and there are no divergences, but
the gauge bosons can be detected and their contribution does not enter
physical observables and it is strongly suppressed in experimental analyses. The violation of IR cancellation theorems leads to soft and collinear divergences at $s\gg m_W^2$ and to double logarithmic enhancements.
For a typical four fermion process the enhancement is \cite{Denner:2000jv}
\begin{equation}
-\f{\alpha}{\pi \sin^2\theta_W} \ln^2\f{s}{m_W^2}
\end{equation}
which at $\sqrt{s}=1$ TeV amounts to about $-25\%$. This implies that EW corrections are increasingly
important at high energies and that their inclusion is crucial in any search at the TeV scale.
This is nicely shown in Fig.~\ref{fig:ew} (from Ref.~\cite{Kallweit:2014xda}) where the impact of NLO
EW corrections on $W+1$ jet events
is compared to the effect of NLO QCD only. At large $p_T$ of the $W$ boson the impact of
the Sudakov effects discussed above is clearly visible, and the EW corrections become more and more important. The effect on the $p_T$ of the leading jet result in an unphysical increase of the cross section at high-$p_T$, due to the large impact of QCD radiation. Applying a cut on the azimuthal separation of the jets reduces the effect of the $W+2$ jet contribution and leads to a more consistent behavior
also in the $p_T$ distribution of the leading jet.

Following the automation of NLO QCD calculations discussed in Sect.~\ref{sec:nlo}, work is ongoing by different groups
to automate the inclusion of EW corrections. 
Besides $W$+multijets \cite{Kallweit:2014xda}, first results are appearing also on vector boson + 2 jets \cite{Denner:2014ina,Chiesa:2015mya} and $t{\bar t}$ + vector and Higgs boson \cite{Frixione:2015zaa}.

\section{Summary}

The first run of the LHC can be considered a triumph for the SM. Besides finding no evidence for
new physics, the last missing ingredient of the theory, the Higgs boson, has been discovered.
The Run 2 has just started and will offer us the possibility to further scrutinize the Higgs sector
and to continue looking for BSM signals.
While eagerly waiting for new data, accurate theoretical predictions
for SM processes are essential to further sharpen our picture on the Higgs boson and to
control SM backgrounds. In the last decade we have witnessed a revolution in NLO QCD calculations, which have now reached a high level of automation. Parallely, work is being devoted to extend this level of
automation also to EW corrections, which are particularly relevant for the high-energy frontier.
When precision plays a role, NNLO QCD corrections become important, and, in particular the possibility
to count on NNLO calculations tailored on the phase space regions in which the measurements are actually done. The past year has seen enormous progress in this direction with many new calculations being completed.
For a key process like Higgs production in gluon fusion, even N$^3$LO corrections have been recently obtained. As is well known, realistic simulations require MC event generators, which offer a complete description
of the hadronic collision. The progress in the NNLO direction is accompanied by first examples of NNLO matched simulations, which will be very useful for Run 2 analysis. In summary, it is fair to say that
the theory community is trying to catch up with the challenges from the LHC, to make the best use of the high-quality data that the experiments will deliver.
\vskip 0.5cm
\noindent {\bf Acknowledgements.} I would like to thank the organizers for making EPS 2015 a lively and enjoyable conference. I am grateful to Stefano Catani, Daniel de Florian, Stefano Forte, Stefano Frixione, Thomas Gehrmann, Stefano Pozzorini, Gavin Salam and Marek Sch\"onherr for useful discussions on the topics presented here. This contribution is dedicated to the memory of Guido Altarelli, who was awarded the EPS prize at this conference and passed away shortly after.

\end{document}